\documentstyle[aps,prl,twocolumn,graphicx]{revtex}

\begin{document}

\draft
\title{Entropic Analysis of non-Stationary Sequences}
\author{Michele Virgilio$^{1}$, 
   and Paolo Grigolini$^{1,2,3}$.}
  \address{$^{1}$Dipartimento di Fisica dell'Universit\`a di Pisa and
   INFM, Via Buonarrotti, 56127 Pisa, Italy}
 \address{$^{2}$Center for Nonlinear Science, University of North Texas,
   P.O. Box 311427, Denton, Texas 76203-1427 }
  \address{$^{3}$Istituto di Biofisica CNR, Area della Ricerca di Pisa,
   Via Alfieri 1, San Cataldo 56010 Ghezzano-Pisa, Italy}
   \date{\today}
   \maketitle

   \begin{abstract}
  The aim of this paper is to shed light on the analysis of  non-stationary time series by means of the method of diffusion entropy. For this purpose,  we first study the case when infinitely many time series, as different realizations of  the same dynamic process, are available, so as to adopt  the Gibbs ensemble perspective. We solve the problem of establishing under which conditions scaling emerges from within this perspective. Then, we study the more challenging problem of creating a diffusion process from only one single (non-stationary) time series. The conversion of this single sequence into many diffusional trajectories  is equivalent to creating a non-Gibbsian
ensemble. However, adopting a probabilistic approach to evaluate  the contribution of any system of this non-Gibbsian ensemble, and using for it the theoretical Gibbsian prescription of the earlier case, we find a recipe that  fits accurately the numerical results. With the help of this recipe we show that  non-stationary time series produce either anomalous scaling with ordinary statistics or ordinary scaling with anomalous statistics. From this  recipe we also derive an attractive way to explain the entropy time evolution, as resulting from two distinct uncertainty sources, the missing information on trajectory initial condition and the randomness  induced lack of control of trajectory time evolution.

   \end{abstract}
   \pacs{89.75.D ,05.70.L,02.30.L ,95.75.W}
   \vspace{0.5cm}
   
   \section{Introduction}
  One of the paradigms of the Science of Complexity is that complex systems, of biological, physical, economical and sociological interest, are characterized by long-range correlations. Thus, a time series obtained by recording data of this kind is expected to reflect
these long-range correlations. How to detect these correlations in practice? One the pioneer papers in this field \cite{dfa}  rests on the study of diffusion process
generated by the time series, whose numbers are imagined to be diffusion generating fluctuations. 
Thus, the detection of these correlations can be made in an indirect way by observing the deviation of the resulting diffusion process
from ordinary Brownian diffusion. 

To set this important issue on a solid ground, let us refer ourselves to the scaling property of the probability distribution function (pdf) $p(x,t)$.
This means that we are assuming that the time series is transformed into many independent, one-dimensional,  diffusion trajectories $x(t)$, and that $p(x,t)$ is the probability density of finding the trajectory in  position $x$ at a given time $t$.
Scaling means that
\begin{equation}
p(x,t) = \frac{1}{t^{\delta}} F(\frac{x}{t^{\delta}}).
\label{scalingdefinition}
\end{equation}
The theoretical foundation of this property is given by the Central Limit Theorem (CLT)\cite{ma}  and by the Generalized Central Limit Theorem (GCLT) \cite{gnedenko}. In short, these two theorems refer to the case when the variable $x$ is the sum of many uncorrelated variables $\xi_{i}$. This is the case when the time series is a sequence of independent fluctuations occurring at the integer times $i$, and $x$ is the sum of the fluctuations ranging from $i = j$ to $i = j+l$, with $l$ playing the role of an integer time that in the limiting case $l \gg 1$ is identified with the continuous variable $t$. Let us now focus our attention on  the fluctuations $\xi_{i}$, and for simplicity, let us
make the assumption that their probability distribution  has  a vanishing mean value. Let us consider, first, the case when the second moment
of distribution is finite. In this case,  the CLT applies. The validity of the CLT is broken when the second moment is divergent. However, in this case, we meet the wonderful opportunity of replacing the CLT with the  GCLT. In the former case $\delta = \frac{1}{2}$, and $F(y) $ is a Gaussian function. In the latter case, $\delta >  \frac{1}{2}$ and $F(y)$ is a L\'{e}vy function, whose asymptotic form is $lim_{|y|\rightarrow \infty} F(y) 
= 1/|y|^{1+1/\delta}$. A third case exists, the popular case of Fractional Brownian Motion (FBM), introduced by Mandelbrot \cite{mandelbrot}. In this case the fluctuations are correlated,  $\delta$ range from $0$ to $1$, and $F(y)$ is a Gaussian function again. The dynamic foundation of this third case is not yet as solid as that of the earlier two cases, and conjectures have been made as to a possible connection between FBM and non-stationary properties \cite{nonstationarybrownianmotion}.

As to the important problem of detecting scaling from the statistical analysis of real time series, the most popular approach is that developed by the authors  of Ref.\cite{dfa}. These authors developed a technique of scaling detection based on the analysis of the variance of the diffusion process generated
by the time series under study. This technique of analysis was termed Detrended Fluctuation Analysis (DFA), and  has been widely applied since then.   In the last few years another method was 
proposed by the authors of Ref. \cite{nicola}. These authors  introduced a technique of scaling detection, based on the evaluation of the entropy of the experimental pdf $p(x,t)$, called Diffusion Entropy Analysis (DEA). We note that in the continuous time limit the Shannon entropy
of the diffusion process reads
\begin{equation}
S(t) = \int_{-\infty}^{+\infty} dx p(x,t) log (p(x,t)).
\label{diffusionentropy}
\end{equation}
By plugging Eq.(\ref{scalingdefinition}) into Eq.(\ref{diffusionentropy}) and using a change of integration variable, we get
\begin{equation}
S(t) = A + \delta log t,
\label{logarithmictime}
\end{equation}
where 
\begin{equation}
A = - \int_{-\infty}^{+\infty} dy F(y) log (F(y)).
\label{theadditiveconstant}
\end{equation}
It is evident that this method can be used to determine $\delta$ without the limitations affecting the DFA. In fact, the DEA works, in principle, even when $F(y)$ does not have a finite second moment, while the DFA, being based on the variance calculation, cannot be applied in the case
of L\'{e}vy statistics.

It is important to stress that both DFA and DEA rest on the creation of a Gibbs ensemble from a single sequence. Thus, to ensure that the systems of this ensemble are identical copies
of the same system, it is necessary to assume a stationary condition. Unfortunately, many complex processes seem to be incompatible with the stationary condition.  An interesting example of non-stationary condition is given, for instance, by the teen birth phenomenon studied in Ref. \cite{nicola}. The time series under study in the paper 
of Ref. \cite{nicola} concern the number of babies born per day to teens in Texas. Since fertility 
has an obvious annual periodicity, and the rate itself of sexual intercourses is holidays dependent, and consequently,
is affected,  too,  by annual periodicities, these kinds of time series are not stationary. The fluctuations of interest 
for  statistical analysis take place around a time dependent mean value dictated by both social and biological deterministic processes.
If the annual periodicities are not properly taken into account, the social process under study might seem to be characterized by memory effects
of surprisingly high intensity. Another case of external modulation of fluctuation intensities is given by the solar flares \cite{debbie}, where the rate of flares is modulated by the 11-years solar cycle.

The main purpose of this paper is to establish which is the consequence of addressing the study of a non-stationary time series with a method originally designed for stationary time series. The paper of Ref.\cite{hu} addresses this important issue using DFA. This paper rests on the use of DEA.

  \section{Gibbs ensemble}
We notice that the statistical analysis of time series which reflect rules changing with time meets with a big conceptual difficulty. This has to do with the fact that a statistical treatment in physics is usually done with the picture of Gibbs in mind. We imagine that the system under study can ideally be repeated a virtually infinite number of times. The copies are exactly identical to the original, and one copy differs from another only for different initial condition. In the case where the system under study undergoes the influence of an environment, expressed under the form of a fluctuation, there is not even the need to use different initial conditions. Moving from one copy to another, the time evolution of the environmental fluctuation changes,  in spite of the constraint that the statistical property of the environmental fluctuation remains unchanged.
 The statistical treatment consists of averaging on this set of virtually infinite copies. If the system under study is, partially or totally, random,
an even slight difference on initial  conditions will create deep differences at later times. The same condition applies to the case of a stochastic system, even if the initial conditions can be chosen to be exactly identical in this case. The observation of the resulting trajectory spreading makes it possible to determine in practice the corresponding pdf.
In the specific case of a time series, we do not have infinitely many identical copies of the same system. However, if the time series is stationary, we can create these copies with the method of the moving window. We move a window of size $l$ along the sequence and all the values 
spanned by the window located at a given position are regarded as values, random, partially random, or stochastic, generating a diffusion trajectory of time length equal to $1$. This cannot be done in the non-stationary case, since moving the window from an old position, with respect to sequence reference frame, to a new one, is equivalent to considering a copy that it is not equivalent to the original copy. We are aware of this difficulty. Nevertheless, in Section 3  we shall study non-stationary time series pretending they are a stationary, and we shall try to see if the resulting effects can give us information on the non-stationary nature of the process. To deepen our understanding of those effects, it is convenient  to study first artificial non-stationary time series. We create a virtually infinite number of copies and we adopt the Gibbs point of view. This is the subject of this section. 

The statistical analysis of time series of the earlier work (see, for instance, Ref. \cite{barbi} as well as Ref.\cite{dfa}) is based on turning a time series into a diffusion process.  The time series $\{\xi_{i}\}$ is a sequence of real numbers that are interpreted as fluctuations. The sum of $l$ of these fluctuations is called $x(l)$ and can be considered as the position of a walker at time $l$. If the sequence is large enough we can consider values of $l$ so large as to interpret $l$ as a continuous time variable that we denote by a different symbol, $t$, to emphasize the adoption of the continuous time approximation. This means that we study the equation of motion
\begin{equation}
\frac{dx}{dt} =  \xi(t).
\label{continuoustimeapproximation}
\end{equation}

To mimic the effect of using a non-stationary time series, we write the fluctuation $\xi(t)$
as follows
\begin{equation}
\xi(t) = g(t)\xi^{0}(t) + h(t),
\label{nonstationaryfromstationary}
\end{equation}
where $g(t)$ and $h(t)$ are  regular functions of time and $\xi^{0}(t)$ a memoryless stochastic process defined by
\begin{equation}
<\xi^{0}(t_{1}) \xi^{0}(t_{2})> =  \delta (|t_{1}- t_{2}|).
\label{whitenoise}
\end{equation}
 The function $g(t)$ serves the purpose of making non stationary the noisy portion of the time series under study and the term $h(t)$ is inspired to the teen birth phenomenon\cite{nicola}. As mentioned in Section 1, fertility has annual periodicity, and its influence has been mimicked by a term $h(t)$, with a sinuisoidal form.

Let us make the change of variables
\begin{equation}
y(t) \equiv \frac{x(t) -\int_{0}^{t}h(t')dt'}{g(t)} .
\label{changeofvariable}
\end{equation}
and let us consider the stochastic process described by
\begin{equation}
\frac{dx}{dt} = g(t) \xi^{(0)}(t)+h(t).
\label{prototype}
\end{equation}
As a consequence of this change of variable, Eq.(\ref{continuoustimeapproximation} ) with $\xi(t)$ defined by Eq.(\ref{nonstationaryfromstationary}), becomes
\begin{equation}
\label{newequation}
\dot y(t) = - \frac{\dot g(t)}{g(t)} y(t) + \xi^{(0)}(t).
\end{equation}
The equation for the probability distribution $p(x,t)$, corresponding to this stochastic equation, is well known to be
\begin{equation}
\frac{\partial p(y,t)}{\partial t}   = ( \frac{\dot  g(t)}{g(t)} \frac{\partial}{\partial y} y + D \frac{\partial^{2}}{\partial^{2} y}) p(y,t).
\label{fokkerplanck}
\end{equation}

Let us look for a solution of Eq.(\ref{fokkerplanck}) with the form:
\begin{equation}
p(y,t) = \frac{1} {[2 \pi \sigma_{y}^{2}(t)]^{1/2}} e^{- \frac{y^{2}}{2\sigma_{y}^{2}(t) }}.
\label{heuristicform}
\end{equation}
By plugging Eq. (\ref{heuristicform}) into Eq. (\ref{fokkerplanck}) we obtain, after some easy algebra,
\begin{equation}
\label{aftersomealgebra}
\frac{d \sigma_{y}^{2}(t)}{dt} + 2 \frac{\dot g(t)}{g(t)} \sigma_{y}^{2}(t)  - 2 D=0.
\end{equation}
This means that Eq. (\ref{heuristicform}) is a solution of Eq.(\ref{fokkerplanck}) provided that the variance
$\sigma_{y}^{2}(t)$ is a solution of Eq.(\ref{aftersomealgebra}). It is straigthforward to prove that
\begin{equation}
\label{variancetimeevolution}
\sigma_{y}^{2} (t) = \frac{2D}{g^{2}(t)} \int_{0}^{t} g^{2}(t')dt'.
\end{equation}
We have selected a solution fitting the condition $\sigma_{y}(0) = 0$, if $g(0) \neq 0$, which corresponds to the way we carry out the statistical analysis of time series \cite{nicola,barbi}. This is so because all diffusion trajectories derived from the adoption of mobile windows are assumed to start from $x = 0$ at $t = 0$. Due to Eq.(\ref{changeofvariable}) this means $y = 0$ at $t = 0$. 
It is straightforward to prove that the probability distribution $p(x,t)$ is then given by
\begin{equation}
\label{pdfofx}
p(x,t) = \frac{1} {[2 \pi \sigma_{x}^{2}(t)]^{1/2}} e^{- \frac{(x - x_{0}(t))^{2}}{2\sigma_{x}^{2}(t) }},
\end{equation}
with
\begin{equation}
\label{sigmaofx}
\sigma_{x}^{2} (t) \equiv g^{2}(t) \sigma_{y}^{2} (t)
\end{equation}
and
\begin{equation}
x_{0}(t) \equiv \int_{0}^{t} h(t')dt'.
\label{drift}
\end{equation}

We are now ready to answer the question as to whether the scaling condition of Eq.(\ref{scalingdefinition}) is compatible with an out of equilibrium condition. It is evident that the conditions to fulfill are:
\begin{equation}
\label{generalcondition}
\sigma_{x}^{2}(t) = 2 D  \int_{0}^{t} g^{2}(t')dt'  = a t^{2 \delta}
\end{equation}
and
\begin{equation}
\label{additional}
\frac{(x-x_{0}(t))^{2}}{2\sigma_{x}^{2}(t)} = \frac{(x - \int_{0}^{t}h(t')dt')^{2}}{2at^{2\delta}} = M(\frac{x}{t^{\delta}})
\end{equation}
By differentiating Eq.(\ref{generalcondition}) with respect to time, we get
\begin{equation}
\label{formofg}
g(t) = (\frac{a \delta}{D})^{1/2} t^{\delta - 1/2}.
\end{equation}
This leads us to the form
\begin{equation}
\label{generalform}
g(t) = b t^{\beta},
\end{equation}
with
\begin{equation}
\label{range}
-\frac{1}{2} < \beta < \infty.
\end{equation}
By expanding the square appearing in Eq. (\ref{additional}), we find for the function $h(t)$ the following form
\begin{equation}
h(t) = c t^{\beta - 1/2}.
\label{ht}
\end{equation}
Thus we conclude that the forms of Eq.(\ref{generalform}) and Eq.(\ref{ht}) are compatible with the scaling 
condition of Eq.(\ref{scalingdefinition}) with the scaling parameter $\delta$ given by
\begin{equation}
\label{scalingfrombeta}
\delta = \beta + \frac{1}{2}.
\end{equation}
We note that the linear case, with $\beta = 1$, yields $\delta = 3/2$, the value obtained as inertial effects in Refs.\cite{masoliver0,masoliver1,masoliver2}.

   \section{Single sequences}
We are now in a position to address the intriguing problem of non-stationary time series when only one time series is available. We do not limit ourselves to applying the DE. We also interpret theoretically the numerical result stemming from the DE. It is the proper time here to concisely review how a single sequence is converted into many trajectories.  Let us imagine that we have the sequence $\{\xi_{i}\}$. From a theoretical point of view, the ideal condition to use would be that the sequence is infinite. Unfortunately, this condition is not 
fulfilled by real sequences, and, for obvious computational limitations, is not fulfilled by artificial sequences either. Thus, we explicitly take this condition into account, and we denote the sequence finite size by $l_{seq}$. We assume that $l_{seq}$, although finite, is large enough as to ensure a fair numerical evaluation of the statistical properties under study.. We denote the maximum temporal length of the trajectories  equal to $l_{max}$. Thus, to make predictions concerning diffusion occurring at time $t = l_{max}$ we can use up to $l_{seq}- l_{max} + 1$ distinct trajectories. These trajectories are not independent of one another.  In fact, the discrete time derivative of the $s$-th trajectory is defined by
\begin{equation}
\label{s-th}
x^{(s)}(t+1) - x^{(s)}(t) = \xi_{s+t}.
\end{equation}
This means the  $s$-th  trajectory and the $(s+1)$-th trajectory, both of lenght $l$,  will have $l-1$ numbers in common, and that only if $|s-s'| > l-1$ the two trajectories are totally independent of one another. It is evident that if we adopted the criterion of non-overlapping windows, as the authors of Ref.\cite{dfa} do, we would get the benefit of using trajectories totally independent of one another without meeting the risk of introducing correlations that do not exist. On the other hand, the number of trajectories to use would be much smaller, and the resulting statistics would be poorer. Thus, we decided to check if the method of overlapping windows yields spurious effects or not. The adoption of artificial sequences makes it possible
for us to settle this important issue. In fact, we did the calculation using the method of overlapping windows and we got a given result. We did again the same calculation with the same number of trajectories, these being totally independent the ones of the others. This means that we adopt the same  stochastic prescription, but we run it several times (one run for each trajectory) so as to create really independent trajectories. The new result turned out to coincide with the earlier, thereby leading us to conclude that the method of overlapping windows can be safely applied. This is in fact the method that we adopt throughout. 

The key formula for the theoretical treatment of the numerical results of this Section is given by
\begin{equation}
p(x,t) = \frac{\sum_{s=1}^{l_{seq} - l_{max} +1} p^{(s)}(x,t)} {l_{seq} - l_{max} + 1}.
\label{keyformula}
\end{equation}
The meaning of this formula is the following. We make the conjecture that the probabilistic contribution of the s-th trajectory to the pdf resulting from this method of numerical observation is the same as that stemming from  an arbitrarily large number of trajectories corresponding to the same physical condition. We do not have a rigorous proof for this theoretical expression, but the extremely good agreement 
between the theoretical prediction itself and the numerical observation. On the other hand, once this theoretical prediction is accepted, the physical interpretation of the numerical results become easier, and as we shall see, we can use this formula to account for two distinct sources of entropy increase, the trajectory randomness and the uncertainty on the initial conditions.

\subsection{Linear increase of noise intensity}

Let us consider first the case where the noise intensity increases linearly with time,

\begin{equation}
\xi(t) = \xi^{(0)} (t)(a + b t) .
\label{linearincrease}
\end{equation}
The sequence corresponding to the $s$-th window, becomes
\begin{equation}
\label{s-thsequence}
\xi^{(s)}(t) = \xi^{(0)}(t + s)(a + bs + bt).
\end{equation}
The $s$-th diffusion trajectory obeys the equation of motion
\begin{equation}
\dot x^{(s)} = \xi^{(0)}(t + s)(a + bs + bt),
\label{s-thdiffusion}
\end{equation}
which is equivalent to
\begin{equation}
\dot x^{(s)}(t) = \xi^{(0)}(t)(a + bs + bt),
\label{equivalents-thdiffusion}
\end{equation}
due to the assumption that the noise $\xi^{(0)}$ is memoryless.
This means that we can refer ourselves to Eq.(\ref{nonstationaryfromstationary}) with
\begin{equation}
g(t) \equiv a + bs + bt, \quad h(t)=0.
\label{effectiveg}
\end{equation}
We plug $g(t)$ of Eq.(\ref{effectiveg}) into Eq.(\ref{variancetimeevolution}). This can be easily integrated and yields
\begin{equation}
\label{sigmainthelinearcase}
\sigma_{x}^2(s, t)  = \frac{2D}{3b} [(a+bs+bt)^{3} - (a+bs)^{3}]  .
\end{equation}
Eq.(\ref{keyformula}) becomes
\begin{equation}
p(x,t) = \frac{1}{(2 \pi)^{1/2} N} \sum_{s=1}^{N} \frac{e^{-\frac{x^{2}}{2 \sigma_{x}^2(s, t)}}}{\sigma_{x}(s,t)} ,
\label{sumofgaussians}
\end{equation}
where
\begin{equation}
N \equiv l_{seq} - l_{max} +1.
\label{definitionofn}
\end{equation}
For $N \rightarrow \infty$, the sum depends on values of $s$ so large as to make $\sigma_{x}^2(s, t) $ very close to $2D(a+bs)^{2}t$. This makes it is easy to assess that $p(x,t)$ becomes the sum of distributions rescaling with $\delta = 1/2$. Thus also $p(x,t)$ rescales with $\delta = 1/2$. This is confirmed by the numerical result illustrated in Fig. 1.

It is remarkable that the shape of the resulting distribution (see Fig. 2) looks quite different from  that of a Gaussian function, as made even more evident from the adoption of the logarithmic scale for the ordinate axis (see Fig. 3). The resulting distribution is very well approximated by the fitting function
\begin{equation}
p_{fit}(x,t)  = \frac{k(t)}{2} e^{- k(t)|x|} .
\label{exponentialform}
\end{equation}
It is remarkable that an exponential form of this kind has been found by Metzler and Klafter \cite{metzlerklafter}. Apparently, the case studied by Metzler and Klafter is quite different from the one illustrated in Figs. 2 and 3. The authors of Ref. \cite{metzlerklafter} find that the exponential structure is associated to the subdiffusional scaling $\delta < 1/2$. Here the scaling, as shown in Fig. 3, is the normal scaling $\delta = 1/2$. However, the similarity might not be accidental. The case discussed by Metzler and Klafter refers to a random walk with long rests and a waiting time, between a jump and the next, given by a  distribution whose asymptotic limit is the same as that of
\begin{equation}
\psi(\tau) = (\mu-1)\frac{T^{\mu-1}}{(T+ \tau)^{\mu}}.
\label{typicalwaitingtimedistribution}
\end{equation}
Sequences of time with this distribution has been studied recently from the point of view of entropy production\cite{massi} . It has been shown, in agreement with the earlier work of Gaspard and Wang\cite{gaspardandwang}, that the entropy production per unit of time is proportional to 
$t^{\mu -2}$. This means that the rate of entropy production is not constant and that for $t \rightarrow \infty$ it tends to vanish. This is a form of non-stationarity that is apparently different from the one studied in this paper. However, in both cases a scaling distribution emerges, and in both cases a striking deviation from the Gaussian form is exhibited. We have also to mention that if we adopt the $\epsilon$- entropy prescription \cite{wanggaspard} to analyze our non-stationary time series, this form of entropy, being proportional to the square of the noise intensity, tends to increase with time. Thus, from the point of view of entropy production
 both sequences are non-stationary, and we wonder if the similarity between the exponential distribution of Fig. 3  and the exponential distributions of Metzler and Klafter is more than accidental.

\subsection{Time series  with noise intensity sinuisoidally dependent on time }
Let us consider the case where
\begin{equation}
g(t) = a cos (\omega t), \quad h(t)=0.
\label{sinusoidaloscilations}
\end{equation}
Adapting the approach of Section 3 A to this case, we get
\begin{equation}
\label{sigmainthefirstsinuisodalcase}
\sigma_{x}^2(s, t)  = a^{2} D(t + \frac{sin 2\omega (t+s) -sin 2\omega s}{2 \omega}).
\end{equation}
It is evident that in the asymptotic time limit the linear term becomes more important than the
harmonic term, which, consequently, can be neglected, thereby forcing the contributions to the distribution
with different $s$, see Eq. (\ref{keyformula}), to become identical to
\begin{equation}
\label{identicalcontributions}
p(x,t) = \frac{1}{[2 \pi \sigma^{2}(t)]^{1/2}} e^{- \frac{x^{2}}{ 2 \sigma^{2}(t)}},
\end{equation}
with
\begin{equation}
\label{standarddiffusion}
\sigma^{2}(t) = a^2 Dt,
\end{equation}
which yields for the DE the following analytical expression
\begin{equation}
\label{sinusoidal1}
S(t) = \frac{1}{2} log t + \frac{1}{2} log (2 \pi e D a^{2}).
\end{equation}
Fig. 4 compares this analytical prediction to the numerical result and shows that, as expected, the theoretical prediction departs from the numerical result in the time region $t < 1/\omega$. 

\subsection{Time series with a sinuisodal deterministic part}
Let us study the case where $g(t) = a$ and $h(t) = b cos \omega t$. In this case, according to the theory of Section 2, for the contribution corresponding to the $s$-th window,  we must specify the form of $x_{0}(t)$ as well as that of $\sigma_{x}(t)$. 
With the same method of Section 3A we get:
\begin{equation}
\sigma_{x}^2(s, t)  = 2 D a^{2} t.
\end{equation}
and
\begin{equation}
\label{driftterm}
x_{0}(s,t) = \int_{0}^{t} h(t'+s)dt'.
\end{equation}
Using $h(t) = b cos \omega t$, with a straigtforward integration, we write
\begin{equation}
\label{driftterm2}
x_{0}(s,t) = I(t) sin (\omega s + \Phi(t)),
\end{equation}
with
\begin{equation}
\label{driftterm3}
I(t) =  \frac{b}{\omega} [ 2(1 - cos \omega t)]^{1/2}.
\end{equation}
The key formula of Eq.(\ref{keyformula}) in this case takes the form
\begin{equation}
\label{specialcaseofkeyformula}
p(x,t) = \frac{1}{N(2 \pi a^{2}t)^{1/2}} \int_{1}^{N} e^{-\frac{(x-x_{0}(s,t))^{2}}{2 a^{2}t}}ds
\end{equation}

In the long-time limit $t >> \frac{b^{2}}{\omega^{2} a^{2} D}$, the contribution of $x_{0}(s,t) $
becomes negligible and we recover  Eq. (\ref{identicalcontributions}) with $\sigma^{2}(t) = 2 Dt$. However, in this case the regime of transition to scaling is much more interesting than that of Section 3 B. At early times
we see a process of entropy increase that might be easily mistaken for a scaling regime with anomalous scaling $\delta > 0.5$. This is caused
by the deterministic process $h(t)$. The adoption of the method of moving windows implies the adoption of many trajectories, all being the solution of the harmonic equation $\frac{d^{2}}{dt^{2}}x  + \omega^{2} x = 0$, with different initial condition. This means that in this case the entropy increase does not depend on the random nature of the time series, but rather on the uncertainty on initial conditions originated by the moving window method. It is worth remarking that all these trajectories are given the same initial position, $x = 0$, but different velocities. The uncertaintly on this initial condition makes the entropy increase. However, all these trajectories at the end of the period $T = 2 \pi/\omega$ must go back to the initial position, $x = 0$, thereby forcing the entropy, too, to regress to the initial vanishing value. This regression is complete if  only the deterministic component is present  $(a = 0)$. This behavior is repeated infinitely many times (see Fig. 5).  It is also remarkable that the presence of an even very small stochastic component provokes  an abrupt transition to a completely different behavior, where the strength of the oscillations, which are a signature of the presence of a deterministic term, tends very slowly to zero.

Fig. 5  must be considered as being  one of the main results of this paper. We have found that also in the case of a single sequence the analysis made by means of the DEA yields a scaling larger than that of normal diffusion, as an effect of the non-stationary bias, yielding the contribution $x_{0}(t)$. The agreement between Eq. (\ref{specialcaseofkeyformula}) and the numerical result is so good as to
give a compelling support to Eq. (\ref{specialcaseofkeyformula})  and, thus,  to Eq.(\ref{keyformula}), from which Eq.(\ref{specialcaseofkeyformula}) is derived. Eq.(\ref{specialcaseofkeyformula}), on the other hand, serves very well the purpose of illustrating 
the joint action of the two sources of entropy increase, the trajectory randomness and the uncertainty on initial conditions.

\section{conclusions}
The first interesting result of this paper is that concerning the dynamic derivation of a diffusion process overlapping the FBM. The FBM is a very popular generalization of ordinary Brownian motion, defined by Eq. (\ref{scalingdefinition}), with $\delta =1/2$.  the distribution $F$ being Gaussian. The FBM keeps the Gaussian form of $F$ and extend $\delta$ to all possible values of $\delta$ within the interval $[0,1]$.
With very simple arguments we have found that this generalized form of Brownian diffusion can be realized by releasing the stationary assumption, a conclusion reminiscent of that reached in earlier work\cite{nonstationarybrownianmotion}. However, in the non-stationary case considered in this paper, it is possible to go beyond the ballistic limit within which the FBM is confined. Thus, we also establish a contact with the earlier research work of Refs. \cite{masoliver0,masoliver1,masoliver2}. 

These results, as interesting as they are, served the main purpose of 
developing a proper perspective to analyze the diffusion process stemming from a single, non stationary time series.This is a challenging problem that, to the best of our knowledge, has been earlier addressed only in the work of Ref.\cite{hu}. We address this issue using the DE as a method   of scaling detection, and we find that the non-stationary condition generates either anomalous scaling and ordinary statistics or ordinary scaling and anomalous statistics. 
We also prove that the presence of a non-stationary bias yields a process of entropy increase that depends on the uncertainty on initial condition, as well as on trajectory randomness. This is a legitimate source of entropy increase that ,however, might lead to the wrong conclusion that the dynamic process described by the time series under study has memory. 
There are other interesting aspects, such as the deviation from Gaussian statistics and its possible connection with the earlier work of Metzler and Klafter\cite{metzlerklafter} , on one side, and Wang ans Gaspard \cite{wanggaspard}, on the other. This is left as a subject for further investigation. 

\cleardoublepage
\begin{figure}[h]
\begin{center} 
\includegraphics[width=7cm]{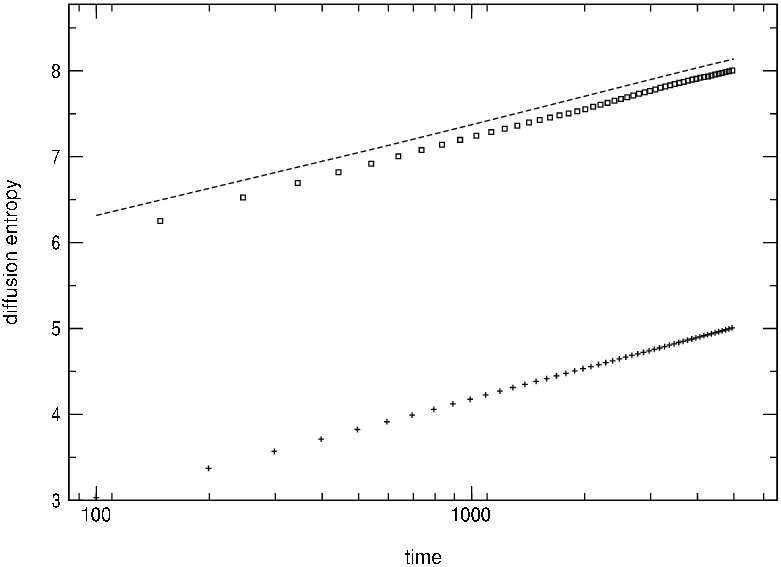} \label{figdelineare} 
\caption{$S(t)$ as a function of time $t$, for the sequence of section III A with  $a=1$, $b=2\cdot 10^{-5}$, $N=10^6$ ($\Box$). The dashed line is the theoretical prediction obtained from eq. (\ref{sumofgaussians}). (+) is the entropy for $a=1$, $b=0$, $N=10^6$ i.e. for the ordinary brownian diffusion.} 
\end{center}
\end{figure} 

\begin{figure}[h] 
\begin{center} \label{figpdflineare}
\includegraphics[width=7cm]{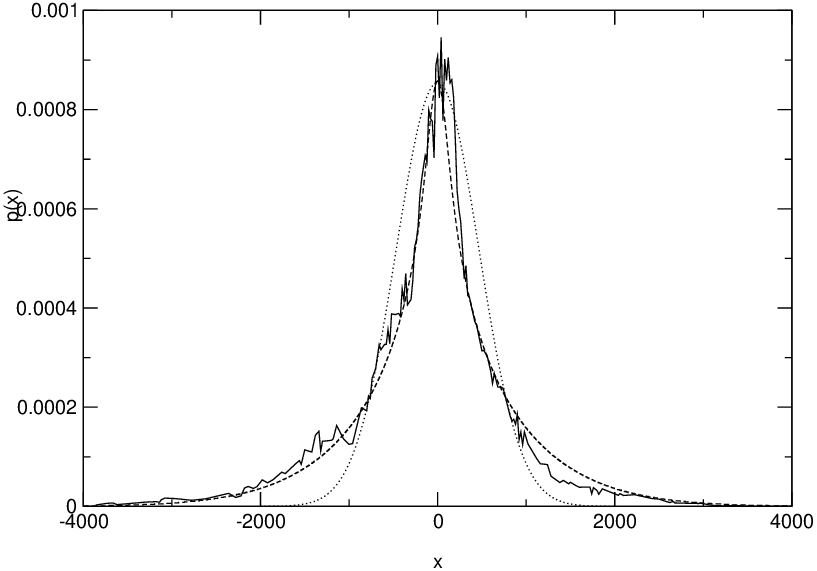} 
\caption{$p(x,t)$ for the set of figure \ref{figdelineare} calculated at $t=5\cdot 10^3$ (solid line). The dashed line stand for the theoretical prediction. The dotted line is a normalizated gaussian.} 
\end{center} 
\end{figure} 

\begin{figure}[h] 
\begin{center}\label{figpdflinearelog} 
\includegraphics[width=7cm]{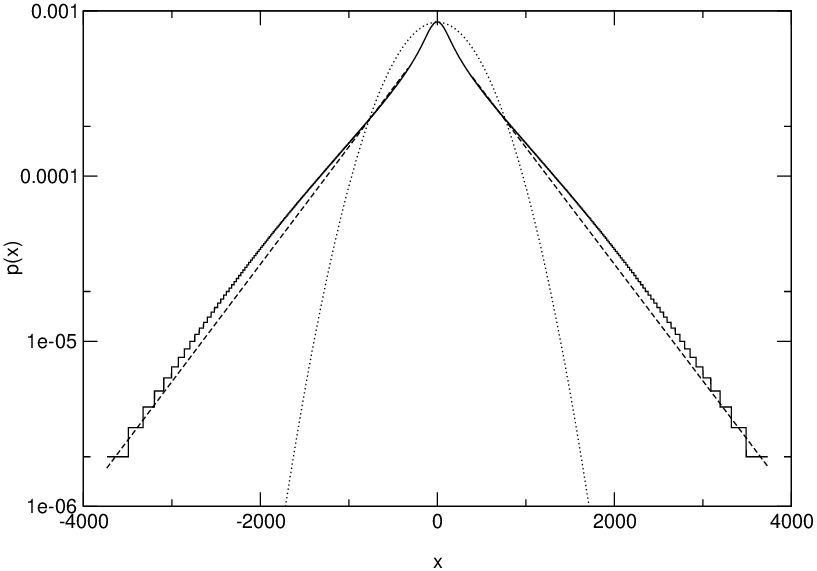} 
\caption{Theoretical prediction  (solid line) and gaussian shape (dotted line) for $p(x,t)$ of figure \ref{figdelineare}. The scale for the ordinate axis is logarithmic.} 
\end{center} 
\end{figure}

\begin{figure}[h]
\begin{center}\label{figde_g_seno}
\includegraphics[width=7cm]{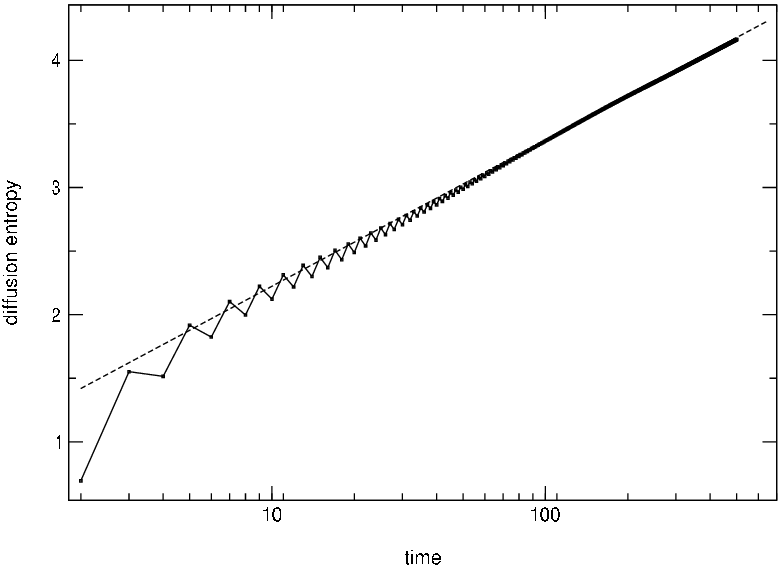}
\caption{$S(t)$ for the sequence of section III B with  $a=1$ (solid line). The dashed line is the theoretical prediction obtained in the asymptotic limit from eq. (\ref{identicalcontributions}).}
\end{center}
\end{figure} 

\begin{figure}[h]
\begin{center}\label{figde_h_seno}
\includegraphics[width=7cm]{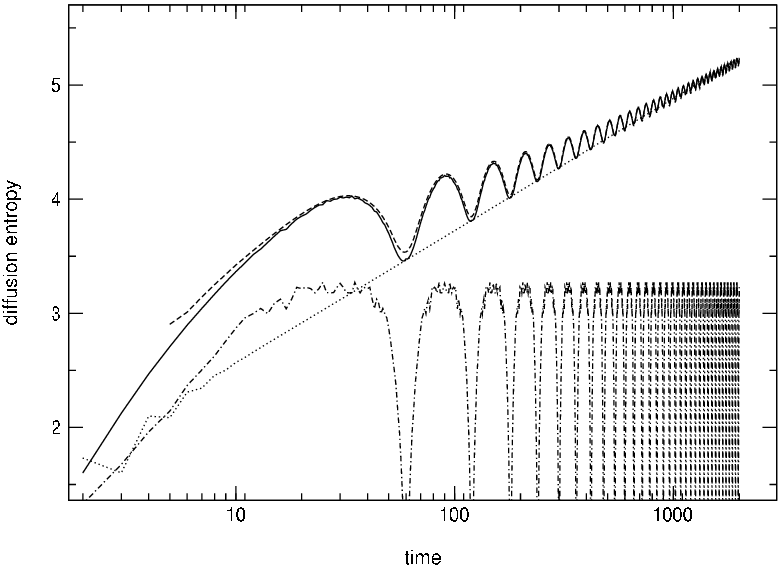}
\caption{$S(t)$ for the set of section III C with $a=b=1$, $T=2\pi/\omega=60$ (solid line). The dashed line is the theoretical prediction obtained from eq. (\ref{specialcaseofkeyformula}). The dotted line is obtained setting $a=1$, $b=0$; the dotted-dashed one setting $a=0$, $b=1$.}
\end{center}
\end{figure}

   \end{document}